\documentclass[a4paper,reqno,12pt,draft]{article}
\usepackage{amssymb,euscript,bbold}
\newcommand{\be}{\begin{equation}}
\newcommand{\ee}{\end{equation}}
\newcommand{\ba}{\begin{eqnarray}}
\newcommand{\ea}{\end{eqnarray}}

\newcommand{\ft}{\footnote}

\begin{document}
\input{epsf}

\begin{flushright}
RUNHETC-2003-07
\end{flushright}
\begin{flushright}
\end{flushright}
\begin{center}
\Large{\sc Compactification with Flux and Yukawa Hierarchies.}\\
\bigskip
{\sc B.S. Acharya}\ft{bacharya@physics.rutgers.edu}\\
\smallskip\large
{\sf Department of Physics,\\
Rutgers University,\\ 126 Freylinghuysen Road,\\ NJ 08854-0849.}

\end{center}
\bigskip
\begin{center}
{\bf {\sc Abstract}}
\end{center}
Flux compactifications of $M$ theory
on certain $G_2$-manifolds have vacua in which all moduli are
determined. These vacua have discrete parameters - the flux quanta. Small changes in these quanta produce small changes in
the moduli and these can lead to large changes in the
Yukawa couplings between Higgses, quarks and leptons.
We argue that a large number of vacua exist with
a wide variety of Yukawa hierarchies. 
\newpage


$M$ theory vacua in which the seven extra dimensions form a
$G_2$-manifold $X$
admit a low energy description in terms of an ${\cal N}$ $=1$
supergravity theory. If $X$ is {\it smooth}, 
the gauge group is abelian and there are no light charged particles. The theory
has $b_3 (X)$ moduli fields $s^i$ whose values determine
the metric of $X$.\ft{$b_3 (X)$ is the number of linearly independent,
harmonic, rank 3 antisymmetric tensor fields on $X$.
In known examples $b_3(X)$ ranges from $2$ to ${\cal{O}}(10^2)$,
but examples probably exist with much larger numbers of moduli.}

When $X$ develops certain kinds of singularities, more
interesting physics emerges. For example, if $X$ develops
an orbifold singularity along a 3-dimensional submanifold
$Q$, then non-Abelian gauge fields are supported at $Q$ (times
the four dimensional macroscopic spacetime) \cite{bsa1}.
These were further studied in \cite{bsa2,amv,aw} and elsewhere. 

If, in addition, $X$ contains a certain kind of conical
singularity at a point $p$ $\subset Q$, then chiral matter
fields are supported at $p$ (again times 
the four dimensional spacetime)\cite{aw,ew1,ew2}\ft{See
\cite{berg,laz,lust} for some generalisations of 
the results in \cite{ew2}}. 
These are charged under the gauge group supported at $Q$.
Various aspects of the phenomenology of these compactifications in the context of high scale unification have been described in \cite{ew3,ew4}.

Note that in principle, $X$ could have many $Q$'s and hence
many gauge groups and matter representations leading to a
vast number of possibilities for the low energy physics.
This already suggests the existence of {\it many} $M$ theory
vacua with a low energy standard model gauge group and
field content but with presumably different couplings.

Low energy $M$ theory physics is well approximated by d=11 supergravity.
The bose fields of this theory are the metric $g$ and a 3-form
field $C$, with field strength $G$. When $X$ develops the singularities discussed above
the action is supplemented with additional gauge and matter
fields which are localised at the singularities. 

We
recently showed that if there is a $Q$ which admits a complex
flat gauge field whose Chern-Simons invariant is not real
and suitably large, then compactifications with non-zero
$G$-flux exist for which the potential has a minimum in which all
the moduli fields $s^i$ are fixed
\cite{bsa4}\ft{We will assume that such $G_2$-manifolds exist. In our opinion, this is a relatively mild assumption.}.
Whilst this mechanism utilises the gauge fields on $Q$ and its superpartners in an
essential way, it does not require invoking  any dynamics
of the charged matter fields and hence does not require the
low energy theory to contain any realistic particle physics.

This is just as well, because the vacuum in the minimum is
supersymmetric with a {\it negative} cosmological constant $\Lambda_0$. If the singularities of $X$ are such that they
produce a particle physics sector which dynamically breaks
supersymmetry, then there will be an additional contribution
to the cosmological constant, $\Lambda_1$, which we will assume is positive. 

In principle, therefore, these models could have positive
cosmological constants. 

Since we are yet to have a good understanding of the supersymmetry
breaking dynamics this is somehwat difficult to evaluate, as
are many other phenomenological questions. However, we feel
that it is nonetheless worthwhile to consider certain
phenomenological aspects of these models before supersymmetry
breaking. After all, if the supersymmetry breaking is ``soft''
enough, it may be that some of what we learn prior to breaking
it persists when it is broken.
Another rationale for this is simple: since we have found a 
vacuum selection principle, we ought to attempt to put it to
some use\ft{In a cosmological setting we are asserting that
the universe is close to the minimum of the scalar potential. Its not atall clear that this is the right thing to do.}!

Because of our lack of understanding of supersymmetry breaking
this ``phenomenology'' will necessarily be crude. 

For example, as we discussed in \cite{bsa4}, in order to
obtain a realistic cosmological constant, $\Lambda_0$ and
$\Lambda_1$ have to be equal to many orders of magnitude.
Without explaining how this could possibly be, we can try to
learn something by imposing that they have the same order
of magnitude. For example, if $\Lambda_1$ is of order
$(TeV)^4$, then the eleven dimensional Planck scale is
of order $10TeV$ and we would have ``large extra dimensions''
as in \cite{large}.
More generally, one can show using the formulae in \cite{bsa4}
that if $\Lambda_1$ is $m^4$ then $M_p$ and $m$ are equal
to within an order of magnitude for certain choices of flux, 
so the scale at which $\Lambda_1$ is generated is at or around
the eleven dimensional Planck scale. So, in a unification
scenario with ``small extra dimensions'', the particle physics
contribution to the cosmological constant will have to be generated
at a high scale, close to $M_{GUT}$. This already {\it suggests}
a departure from conventional wisdom. Firstly the 
standard lore suggests that the highest scale of supersymmetry
breaking (which occurs in certain
models of gravity mediated supersymmetry breaking) 
is $10^{13}GeV$,
whereas in the high scale unification based on $G_2$-manifolds
the fundamental Planck scale is several orders of magnitude
higher, not one. In reality this observation probably means that
$\Lambda_0$ and $\Lambda_1$ are not of the same order in these
models atall.

However, even if we do not solve the cosmological constant
problem, we may have more success in explaining other
observed features of our universe. In this paper we will
show that the $M$ theory vacua with flux described in \cite{bsa4} naturally predict hierarchies of Yukawa couplings
prior to supersymmetry breaking. In fact, we will see that
by tuning the discrete parameters, one can obtain models with
a wide variety of Yukawa hierarchies. 

\cite{banks} emphasised the fact that couplings depend
on flux quanta from a general point of view. Here we will
be able to be more specific.

Including supersymmetry
breaking dynamics is unlikely, in our opinion, to
change the conclusion that a wide variety of vacua exist
with a wide variety of Yukawa couplings. If so, the
conclusion of this paper will be that there are {\it many}
vacua of $M$ theory which contain the field content
of the standard model at low energies, but with
the incorrect masses and couplings. Moreover {\it some} of these
vacua do have the correct couplings.

The vacua under discussion have many discrete parameters.
For each modulus field we have to specify an integer $N_i$
which labels the amount of $G$-flux in a particular
direction in the cohomology lattice of $X$. Changing the
flux quanta $N_i$ changes the value of the $s^i$ in the
minimum of the potential. So there are actually many different
vacua, with many different values of the moduli.

If $Q$ is the 3-manifold which supports the standard model
gauge group, then the quark, lepton and Higgs fields are supported
at points $p_i$ on $Q$. Because in general, the $p_i$ are
distinct in the extra dimensions, there can be no local interactions between charged matter particles in these vacua.
Instead, nonlocal interactions are generated by Euclidean membrane
instantons\ft{In the smooth case, these have been studied in
\cite{jeff}} whose world volumes are 3-manifolds $\Sigma$ which
pass through the points $p_i$. For example, if $p_1 , p_2$
and $p_3$ support a Higgs field $H$, and charged chiral fermion fields $\psi_1$ and $\psi_2$, then a Yukawa coupling
\be
L \sim \lambda H\psi_1 \psi_2
\ee

can be generated if $X$ contains a 3-manifold $\Sigma$ passing
through the three points. The Yukawa coupling constant 
$\lambda$
will be exponentially small in the volume of $\Sigma$
\be
\lambda \sim e^{-2\pi Vol(\Sigma )}
\ee

In particular, since we expect the Yukawa couplings to be generated
through superpotential interactions, we expect that
$\Sigma$ is a supersymmetric 3-cycle in $X$ (as is $Q$).
This implies that 
\be
Vol(\Sigma) = a_i s^i
\ee
ie is linear in the moduli.  So we see from these simple
observations that small changes in the moduli can produce
large changes in Yukawa couplings.

As we will see, choices of flux quanta
exist in which the hierarchies of Yukawa couplings 
in these vacua are roughly
compatible with the observed Yukawa couplings of the standard
model. Many more vacua could also exist in which the hierarchies of Yukawa couplings are unlikely to be correct.

In \cite{bsa4} we found that in the vacuum, the moduli fields
satisfied the relation
\be
\int_X G \wedge \varphi = \Sigma_i 2\pi N_i s^i = 
-{7 \over 5}c_2
\ee
Where $G$ is the four-form flux and $\varphi$ is the moduli
dependent 3-form $G_2$-structure on $X$. $c_2$ is the imaginary part of the Chern-Simons invariant associated with
the superpotential induced by the gauge fields and superpartners. $c_2$ is a non-zero constant in the vacua
under consideration.

This equation shows that if, say, $N_1$ is increased, $s^1$
must decrease. For simplicity we will assume that each
Yukawa coupling comes from one supersymmetric 3-cycle and
further that the volume of each such 3-cycle is one of the
$s^i$ ie that the 3-cycles which give rise to the Yukawa
couplings actually form part of a basis for the homology
of $X$. These assumptions do not affect the results.

Then the $i$'th Yukawa coupling is of order
\be
\lambda_i \sim exp({1.4 c_2 \over b_3(X) N_i})
\ee
where $b_3(X)$ is the number of independent flux quanta.

Moreover, the ratio of two Yukawa couplings is
\be
{\lambda_i \over \lambda_j} \sim 
exp({1.4 c_2 }{N_j - N_i \over b_3(X) N_i N_j})
\ee

Note that these formulae should be interpreted as giving the
Yukawa couplings at the eleven dimensional Planck scale $M_p$.
In a given model we have to renormalise these couplings
to, say, the electroweak scale in order to compare with
experiment. This really depends upon the details of the
model if the Planck scale is high. However, as we will
show below, choices of flux exist which can easily produce
a large variety of hierarchies of couplings at the Planck scale, so
if a given compactification requires a particular hierarchy,
then 
choices of flux exist which could produce it.

To illustrate these points, let us assume that the effective field
theory is such that the Yukawa couplings of the light fermions
are not renormalised too much between $M_p$ and say $1GeV$.
The Yukawa coupling of the electron is then roughly $10^{-6}$.
So if $\lambda_1$ is this coupling, then ${c_2 \over b_3(X)N_1}$
$\sim -9.8$. If, for example, ${c_2 \over b_3(X)} = 3026$, then
choosing $N_1 = 307$ makes $\lambda_1$ the right order of
magnitude. Then, if $\lambda_2$ is the Yukawa coupling of
the muon, which is about two orders of magnitude larger we
can take a flux vacuum in which $N_2 = 460$.  The difference
between $N_1$ and $N_2$ is of order 150, so by choosing
values of $N_2$ closer to $N_1$ we have a lot of freedom
to tune the ratio of couplings. 
So, if the effective field theory
is such that the required Yukawas at the Planck scale are
much closer than those observed at low energies, this can
easily be achieved. Finally, we should mention that since
the Yukawa coupling of the top quark is so large, that a
fairly large flux might be necessary, 
which produces an order one contribution in $M$ theory.

Some comments are in order. If $c_2$ is lowered, then the gap
between $N_1$ and $N_2$ is lowered, and consequently the range
of possible values of ${\lambda_2 \over \lambda_1}$ decreases.
So, for manifolds with low $c_2$, the number of possible vacua
with the correct masses decreases. This suggests that as
$c_2$ is lowered the predictive power of the theory increases
since the number of models that one has to analyse goes
down. We remind the reader that lowering $c_2$ decreases
the size of the extra dimensions and therefore increases
$M_p$.

But if $c_2$ is large enough we see that before taking into
account any supersymmetry breaking dynamics, many vacua exist
with a wide variety of Yukawa hierarchies.

The vacua with different choices of flux are ``related''
via domain wall nucleation. In this case, if $G_1$ and $G_2$
are two different choices of $G$-flux then the domain
wall connecting them is an $M5$-brane which wraps the
3-manifold $D$ in $X$ whose homology class is dual to the
difference of the two $G$-fluxes:
\be
[D] = G_1 - G_2
\ee

Bousso and Polchinski have argued that since $X$ could have
many 3-cycles that it is likely that vacua exist for which
the value of the cosmologial constant is consistent with
observation \cite{bp}. As we have already mentioned, before taking
into account any supersymmetry breaking dynamics, the cosmological constant is negative, regardless of the flux
quanta. However, it may be that the arguments of \cite{bp}
are valid after supersymmetry breaking.

From our discussion here, we can see that,
even though we have to tune the flux quanta carefully to obtain the correct fermion masses, we only have to tune
a few fluxes, and not hundreds. Therefore, accepting the
the arguments in \cite{bp}\ft{For a further discussion see \cite{banks}.}, 
we might conclude that vacua exist with
the correct cosmological constant {\it and} fermion masses.
However, fine tuning of many more fluxes may be required
to obtain the correct superpartner masses, say and other
couplings of nature. This is difficult to estimate since
we do not understand superymmetry breaking well enough.
But it may turn out that the arguments of \cite{bp}
could be invalidated, since even though universes with the
correct cosmological constant exist, it  could be unlikely
that they would have the correct particle masses and interactions.

\cite{shamit}
describe supersymmetric Type IIB string flux vacua with
negative $\Lambda$ and fixed moduli. Studies of these
vacua are likely to lead to similar sorts of conclusions
to those drawn here.
These and more general questions have been addressed 
recently in \cite{mrd} which takes a more statistical approach to the problem
of ``standard model vacua'' of string and $M$ theory.

\bigskip
\large
\noindent
{\bf {\sf Acknowledgements.}}
\normalsize
We would like to thank the Theorists at MIT for stimulating
the discussion from which this paper was conceived, Thomas Dent, Mike Douglas, Steve Gubser and Greg Moore.


\begin{thebibliography}{10}

\bibitem{bsa1} B.S. Acharya, {\sf M theory, Joyce Orbifolds and Super Yang-Mills,} [arXiv:hep-th/9812205].
\bibitem{bsa2} B.S. Acharya, {\sf On Realising N=1 super Yang-Mills in M theory.} [arXiv:hep-th/0011089];{\sf Confining
Strings from $G_2$-holonomy Spacetimes,}[arXiv:hep-th/0101206]
\bibitem{amv} M. Atiyah, J. Maldacena and C. Vafa, {\sf An M theory flop as a Large N duality,}
[arXiv:hep-th/0011256].
\bibitem{aw} M. Atiyah and E. Witten,
{\sf M theory Dynamics on a Manifold of $G_2$-holonomy,} [arXiv:hep-th/0107177].
\bibitem{ew1} E. Witten, {\sf Anomaly Cancellation on $G_2$-manifolds,}
[arXiv:hep-th/0108165].
\bibitem{ew2} B.S. Acharya and E. Witten,
{\sf Chiral Fermions from Manifolds of $G_2$-holonomy,} [arXiv:hep-th/0109152].
\bibitem{berg} P. Berglund and A. Brandhuber, {\sf Matter from $G_2$-manifolds,}
[arXiv:hep-th/0205184].
\bibitem{laz} L. Angelova and C. Lazaroiu, [arXiv:hep-th/0204249],
[arXiv:hep-th/0205070].
\bibitem{lust} K. Behrndt, G. Dall'Agata, D. Lust, S.Mahapatra, [arXiv:hep-th:0207117].
\bibitem{ew3} E. Witten, {\sf Deconstruction, $G_2$-holonomy and Doublet-Triplet Splitting,}
[arXiv:hep-ph/0201018].
\bibitem{ew4} T. Friedmann and E. Witten,
{\sf Unification Scale, Proton Decay and Manifolds of $G_2$-holonomy,}
[arXiv:hep-th/0211269].
\bibitem{bsa4} B.S. Acharya, {\sf A Moduli fixing Mechanism in
$M$ theory,} hep-th/0212294.
\bibitem{large}N. Arkani-Hamed, S. Dimopoulos and G. Dvali,
{\sf The hierarchy Problem and New Dimensions at a Millimeter,} Phys. Lett. {\bf B429} (1998) 263 
[arXiv:hep-ph/9803315];I. Antionadis, N. Arkani-Hamed,
S. Dimopoulos and G. Dvali, {\sf New Dimensions at a Millimeter to a Fermi and Superstrings at a Tev} 
Phys. Lett. {\bf B436} (1998) 257. 
\bibitem{jeff} J. Harvey and G. Moore, 
{\sf Superpotentials and Membrane Instantons,}
[arXiv:hep-th/9907026].
\bibitem{bp} R. Bousso and J. Polchinski, {Quantization of
Four-Form Flux and Dynamical Relaxation of the Cosmological
Constant,} [arXiv:hep-th/0004134]JHEP 0006 (2000) 006
\bibitem{banks}T. Banks, M. Dine and L. Motl, {On Anthropic Solutions of the Cosmological Constant Problem,} 
JHEP 0101 (2001) 031 [arXiv: hep-th/0007206] 
\bibitem{sugraguts}L. Alvarez-Gaume, J. Polchinksi and M. Wise, {\sf Minimal Supergravity GUTS,} Nucl. Phys. {\bf B221}
(1983) 495; H.P. Nilles, Phys. Rep. {\bf C 110} (1984) 1.
\bibitem{shamit} S. Kachru, R. Kallosh, A. Linde and
S. Trivedi, {\sf de Sitter Vacua in String Theory,} 
[arXiv:hep-th/0301240].
\bibitem{mrd} M.R. Douglas, {\sf The Statistics of String/M theory Vacua,}[arXiv:hep-th/0303194].

\end{thebibliography}
\end{document}